\documentclass[pre,aps,twocolumn,superscriptaddress,noeprint]{revtex4-2}

\usepackage[dvips]{graphicx}
\usepackage{amssymb,amsfonts,amsmath,gensymb}
\usepackage{xcolor}
\usepackage[hidelinks]{hyperref}
\usepackage[capitalise]{cleveref}
\usepackage{siunitx}

\newcommand{\ind}[1]{_\mathrm{#1}}
\newcommand*\diff{\mathop{}\!\mathrm{d}}

\newcommand*{\intd}[1]{\int\!\diff #1\,}

\newcommand{\mum}[0]{\mu_m}
\newcommand{\kB}[0]{k\ind{B}}
\newcommand{\kT}[0]{\kB T}

\newcommand{\mue}[0]{\mu_{\rm eff}}
\newcommand{\fp}[0]{f_{\rm P}}
\newcommand{\rhoe}[0]{\rho_{\rm EOS}}

\newcommand{\mbuoy}[0]{m_{\rm b}}
\newcommand{\etab}[0]{\bar{\eta}}

\newcommand{\muc}[0]{\mu_{\rm c}}

\newcommand{\eg}{e.g.\ }
\newcommand{\ie}{i.e.\ }

\newcommand{\citeau}[1]{\citeauthor{#1}~\cite{#1}}

\begin{document}

\title{Effect of sample height and particle elongation in the sedimentation of colloidal rods}

\author{Tobias Eckert}
\affiliation{Theoretische Physik II, Physikalisches Institut,
  Universit{\"a}t Bayreuth, D-95440 Bayreuth, Germany}

\author{Matthias Schmidt}
\affiliation{Theoretische Physik II, Physikalisches Institut,
  Universit{\"a}t Bayreuth, D-95440 Bayreuth, Germany}

\author{Daniel de las Heras}
\email{delasheras.daniel@gmail.com}
\homepage{www.danieldelasheras.com}
\affiliation{Theoretische Physik II, Physikalisches Institut,
  Universit{\"a}t Bayreuth, D-95440 Bayreuth, Germany}

\begin{abstract}
	We study theoretically the effect of a gravitational field on the equilibrium behaviour of a colloidal suspension of rods with different length-to-width aspect ratios.
	The bulk phases of the system are described with analytical equations of state.
	The gravitational field is then incorporated via sedimentation path theory, which assumes a local equilibrium condition at each altitude of the sample.
	The bulk phenomenology is significantly enriched by the presence of the gravitational field.
	In a suspension of elongated rods with five stable phases in bulk, the gravitational field stabilizes up to fifteen different stacking sequences.
	The sample height has a non-trivial effect on the stable stacking sequence.
	New layers of distinct bulk phases appear either at the top, at the bottom, or simultaneously at the top and the bottom when increasing the sample height at constant colloidal concentration.
	We also study sedimentation in a mass-polydisperse suspension in which all rods have the same shape but different buoyant masses.
\end{abstract}

\date{\today}
\maketitle

\section{Introduction}

Hard particles possess an interaction potential that is infinite if two particles overlap and zero otherwise.
Hard spherocylinders, which are cylinders capped with hemispheres at both ends, are among the most popular hard particle models~\cite{Mederos2014}, partly because computing whether two particles overlap or not is relatively simple, and also because their phase behaviour is rich.
Hard models are suitable candidates to study the phase behaviour of colloidal systems since the interaction between colloidal particles is often short-ranged and primarily repulsive.
As revealed by computer simulations~\cite{Frenkel1988,Veerman1990,McGrother1996,Bolhuis1997} hard spherocylinders can form isotropic, nematic, smectic, and crystalline phases depending on their length-to-width aspect ratio and the overall packing fraction. 
The percolation~\cite{Balberg1984,Bug1985,Schilling2015,Xu2016}, random close packing~\cite{FerreiroCordova2014,Meng2016,Freeman2019}, and random sequential adsorption~\cite{Ricci1994,Ciesla2019} of hard spherocylinders have been subject of intense investigation due to the versatility of the model to describe a wide range of systems ranging from lyotropic liquid crystals to granular particles.
Theoretically, classical density functional theory~\cite{Evans1979} has been widely used to study the phase behaviour of hard spherocylinders via functionals based on smoothed density approximations~\cite{Poniewierski1988,Poniewierski1991}, weighted density approximations~\cite{Somoza1990,Velasco2000}, and fundamental measure theory~\cite{HansenGoos2009,Haertel2010,Wittmann2016}.

Beyond bulk phenomena, several works have focused on inhomogeneous systems of hard spherocylinders.
Interfacial phenomena~\cite{Holyst1988,McMullen1988,Moore1990,Velasco2002,Wittmann2014},
wetting~\cite{Dijkstra2001,Heras2003,Brumby2017}, capillary nematization~\cite{Dijkstra2001,Lagomarsino2003,Heras2004,Brumby2017,Basurto2020} and smectization~\cite{Heras2005,Heras2006} in planar pores, as well as confinement-induced phenomena in different geometries~\cite{Dzubiella2000,Trukhina2008,ViverosMendez2017,Rajendra2023} have been studied with density functional approximations and computer simulations.

The hard spherocylinder model has been also used as a reference system to build up more complex interparticle potentials.
These include spherocylinders with dipolar~\cite{Weis1992,Levesque1993a,Williamson1997,McGROTHER1998,Shelley1999}, Coulombic~\cite{Avendano2008,JimenezSerratos2011} and patchy~\cite{Vacha2011,Zhang2015,Jurasek2017} interactions, active spherocylinders~\cite{Bott2018,Yan2019,Stengele2022,Alaniz2023}, as well as spherocylinders coated with soft layers~\cite{Cuetos2002,CamposVillalobos2021}.
Moreover, the hard spherocylinder model can be a reasonable approximation to the shape and the interaction of real colloidal particles such as natural clay rods~\cite{Zhang2006}, $fd$ virus~\cite{Purdy2005}, rod-like boehmite particles~\cite{Buining1993}, polystyrene ellipsoids~\cite{Shah2012}, silica rods~\cite{Ding2009,Kuijk2012,Abbott2018}, as well as PMMA rods~\cite{Keville1991,Mukhija2011} and ellipsoids~\cite{Roller2020,Roller2021}.

Sedimentation experiments, in which a colloidal suspension is equilibrated under the influence of a gravitational field, are one of the basic tools to investigate phase behaviour in colloidal science.
\citeau{Zhang2006} investigated the isotropic-nematic transition in a polydisperse suspension of natural clay rods in sedimentation.
Polydispersity induces a nematic-nematic phase separation with strong fractionation in the rod length.
\citeau{Kuijk2012} performed sedimentation experiments on silica rods with different aspect ratios and constructed an approximated bulk phase diagram by estimating the packing fractions at different heights.
Beyond isotropic, nematic, and smectic A phases, they found a smectic B phase that preempts the formation of a full crystalline state, likely due to polydispersity and the presence of charges.

The gravitational field can have a strong and far from trivial effect in sedimentation experiments, especially if the gravitational length is smaller or comparable to the height of the vessel~\cite{Eckert2022}, which is often the case in colloidal science.
To correctly extract information about bulk phenomena from sedimentation experiments it is essential to understand the effect of the gravitational field on the suspension.
Not much theoretical and simulation work has been devoted to understand the effects of gravity on a suspension of hard spherocylinders.
\citeau{ViverosMendez2014} studied with Monte Carlo simulations the sedimentation of a neutral mixture of oppositely charged spherocylinders.
\citeau{Savenko2004} studied sedimentation-diffusion-equilibrium in suspensions of hard spherocylinders with length-to-width aspect ratio of $5$ using the macroscopic osmotic equilibrium conditions, and compared the results to Monte Carlo simulations.
Depending on the average packing fraction and the height of the sample they found a rich variety of stacking sequences with up to four layers of different bulk phases (top isotropic followed by nematic, smectic and finally a bottom layer of a crystal phase).

Here, we do a systematic theoretical study of the effect of particle elongation and sample height in the sedimentation of suspensions of colloidal hard spherocylinders in equilibrium.
We use the equations of state (EOS) proposed by~\citeau{Peters2020a} to describe the bulk of the suspensions, and sedimentation path theory~\cite{Heras2012,Heras2013} to incorporate the effect of gravity.
The EOS by~\citeau{Peters2020a} are based on scaled particle theory~\cite{Cotter1977} and extended cell theory~\cite{Graf1999}, and reproduce quantitatively the full phase behaviour of hard spherocylinders for all aspect ratios.
We study sedimentation of monodisperse suspensions with four characteristic values of the aspect ratio that cover the whole range of bulk phase phenomena.
We also investigate the evolution of the stacking sequences upon varying the height of the vessel at constant packing fraction.
By increasing the sample height new layers can appear in the sample either at the top, the bottom, or simultaneously at the top and the bottom of the sample.
Using a recent extension of sedimentation path theory to mass-polydisperse suspensions~\cite{Eckert2022b}, in which the particles differ only in their buoyant mass, we study the interplay between mass-polydispersity and gravity in suspensions near density matching.
Under such conditions it is possible to find stacking sequences with up to seven layers and, in contrast to monodisperse systems, new layers can also appear in the middle of the sample.

\section{Theory}

A full account of the theory for monodisperse and mass-polydisperse colloidal suspensions is given in Ref.~\cite{Eckert2022b}.
Here, we give only a brief summary of the theory.

\subsection{Sedimentation path theory for monodisperse and mass-polydisperse systems}
\begin{figure}
  \centering
  \includegraphics[width=0.95\linewidth]{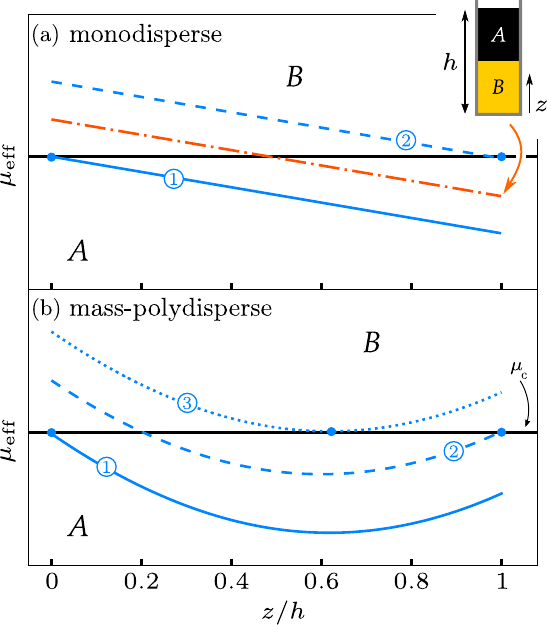}
	\caption{{\bf Sedimentation path theory.} Schematic plots of the effective chemical potential $\mue$ vs the altitude $z$ scaled with the sample height $h$ in sedimentation samples of monodisperse (a) and mass-polydisperse (b) colloidal suspensions.
	The horizontal black line indicates the chemical potential $\muc$ of a bulk transition between two phases labeled $A$ and $B$. 
	Several representative sedimentation paths are plotted.
	The paths are straight lines in monodisperse suspensions (a) and curves in mass-polydisperse suspensions (b).
	The sample corresponding to the orange dash-dotted path is sketched in the inset of (a).
	The blue paths form the sedimentation binodals between two different stacking sequences. 
	An infinitesimal change of any of the blue paths can alter the stacking sequence.
	Represented are paths that end at the bulk coexistence, \ie $\mue(z=0)=\muc$ (blue-solid paths marked with an encircled $1$),
	paths that start at the binodal, \ie $\mue(z=h)=\muc$ (blue-dashed paths marked with an encircled $2$), and
	a path that is tangent to the binodal (blue-dotted path marked with an encircled $3$).
  }
  \label{fig0}
\end{figure}

Sedimentation path theory is based on a local equilibrium approximation to describe sedimentation-diffusion-equilibrium in a colloidal suspension under a gravitational field. 
In monodisperse suspensions, the state of the sample at altitude $z$ (measured from the bottom of the sample) is approximated by a bulk state with an effective chemical potential given by
\begin{equation}
\mue(z)=\mu^0-\mbuoy gz,\;\;\; 0\le z\le h.
\label{eq:mue-mono}
\end{equation}
Here $\mbuoy$ is the buoyant mass of the particles, $g$ is the acceleration of gravity, $h$ is the sample height and $\mu^0$ is a constant offset of the effective chemical potential that can be interpreted as the colloid chemical potential at position $z=0$.
The value of $\mu^0$ determines the total colloidal density of the sample.
The local equilibrium approximation is accurate provided that all correlation lengths are small compared to the gravitational length $\xi=\kT/(\mbuoy g)$, with $\kB$ being the Boltzmann's constant and $T$ being the absolute temperature.

In the plane of $z$ and $\mue$ the effective chemical potential is just the segment of a line, see Fig.~\ref{fig0}(a), known as the sedimentation path~\cite{Heras2012,Heras2013}.
The sedimentation path is directly related to the stacking sequence observed in sedimentation, \ie the sequence of layers of different bulk phases.
An interface appears in the vessel when the sedimentation path crosses a bulk binodal. 
That is, an interface between two bulk phases $A$ and $B$ at position $z_0$ occurs if $\mue(z_0)=\muc$ with $\muc$ the bulk chemical potential of the coexisting phases, see the orange sedimentation path in Fig.~\ref{fig0}(a).

In a mass-polydisperse suspension there is a distribution of colloidal particles that differ only in their buoyant masses.
Such system can be experimentally created with \eg core-shell colloidal particles~\cite{Lu2002,Roller2020} of identical overall size and shape but with internal cores of different sizes.
The interparticle interactions and hence the bulk phenomena (without gravity) are the same as in the corresponding monodisperse suspension.
This greatly simplifies the theoretical treatment while it still highlights the interplay between polydispersity and gravity.

The effective chemical potential of the species of mass $m$ in a gravitational field is in generalization of Eq.~\eqref{eq:mue-mono}  approximated by 
\begin{equation}
  \label{eq:mum}
	\mum(z) = \mum^0 - \mbuoy gz,\;\;\;0\le z\le h.
\end{equation}
Here $m=\mbuoy/m_0$ is a scaled buoyant mass with $\mbuoy$ being the actual buoyant mass and $m_0$ being a reference buoyant mass such that $m$ is dimensionless.
A sensible choice is to relate $m_0$ with either the average buoyant mass or with the standard deviation of the parent distribution, which is the initial distribution of buoyant masses in bulk.
The constant $\mum^0$ sets the overall density of the species with mass $m$ in the suspension.
Using Eq.~\eqref{eq:mum} for the chemical potential of species with mass $m$, one arrives at the following exact expression for the effective chemical potential of the suspension along the sedimentation path~\cite{Eckert2022b}
\begin{equation}
  \label{eq:mu-eff}
  \mue(z) = \kT \ln \left( \intd{m} e^{\beta(\mum^0 - \mbuoy gz)} \right),\;\;\;0\le z\le h,
\end{equation}
where the integral is over the whole range of masses in the parent distribution.
We therefore have successfully mapped a mass-polydisperse system under gravity onto an effective monodisperse system with local chemical potential $\mue(z)$.
The sedimentation path is no longer a straight line, although the \texttt{LogSumExp} structure of Eq.~\eqref{eq:mu-eff} imposes severe restrictions to the possible shapes of the path.
In particular, it follows from Eq.~\eqref{eq:mu-eff} that $\mue(z)$ is a concave function of the altitude $z$ and hence it can have at most one strict minimum.
If only a single value of the buoyant mass is allowed in the parent distribution, then the integral in Eq.~\eqref{eq:mu-eff} collapses to a single buoyant mass and the effective chemical potential reduces to the monodisperse case, Eq.~\eqref{eq:mue-mono}.
That is, the mass-polydisperse system contains the monodisperse system as the limit in which the parent distribution is a delta distribution.

To translate between chemical potentials and densities, we need a bulk equation of state (EOS) in the form of density $\rhoe(\mu)$ as a function of the chemical potential $\mu$, which describes the bulk phase behavior of the system, \ie in absence of gravity.
Then, using the equation of state together with Eqs.~\eqref{eq:mum} and~\eqref{eq:mu-eff} we obtain $\rho_m(z)$, the density profile of the species with mass $m$ and $\rho(z)=\int \!\diff m\,\rho_m(z)$, the overall density profile across all species.

Usually, we set an initial (desired) parent distribution of particles and then find the offsets $\mum^0$ in Eq.~\eqref{eq:mu-eff} that reproduce the parent distribution.
That is, we find the offsets $\mum^0$ such that the overall density of species $m$ in sedimentation, which is $1/h\int_0^h\!\diff z \, \rho_m(z)$, is equal to the density of particles with mass $m$ in the parent distribution.
The offsets are found numerically via a simple iterative procedure~\cite{Eckert2022b}.

\begin{figure*}[t]
  \centering
  \includegraphics[width=1.00\linewidth]{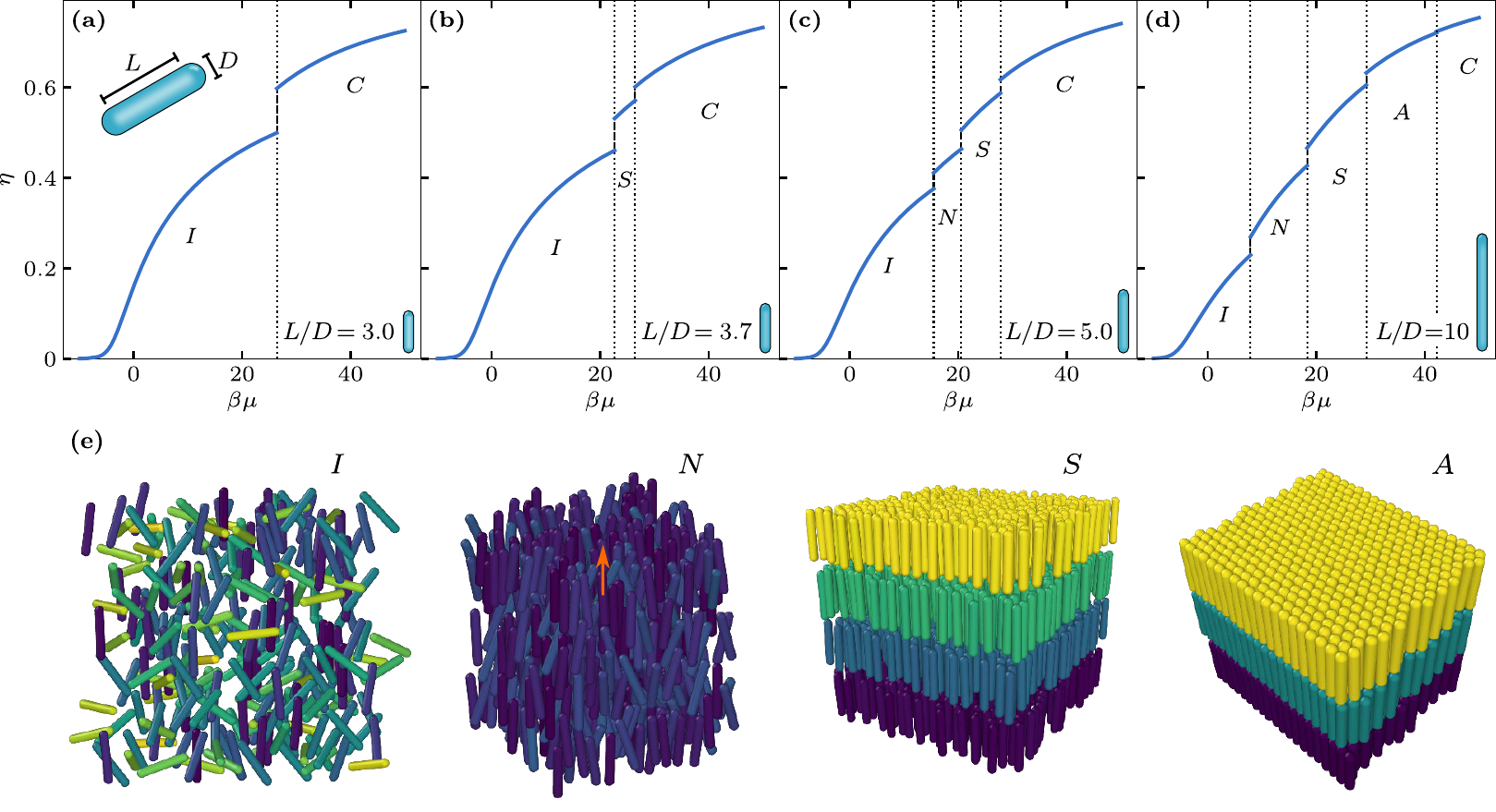}
	\caption{{\bf Bulk phenomena.} Bulk equation of state in the plane of chemical potential $\beta\mu$ and packing fraction $\eta$ for colloidal hard spherocylinders
	according to~\citeau{Peters2020a} for different length-to-width aspect ratios: (a) $L/D$=3.0, (b) $L/D$=3.7, (c) $L/D=5.0$, and (d) $L/D=10.0$.
	The vertical dotted lines indicate the position of the (first order) phase transitions.
	The occurring bulk phases are isotropic $(I)$, nematic $(N)$, smectic-A $(S)$, AAA crystal $(A)$, and ABC crystal $(C)$.
	Panel (e) shows sketches of spherocylinders forming several bulk phases: isotropic state with no positional order and no orientational order, nematic state with no positional order and with orientational order along the director (orange arrow), smectic state with orientational order and positional order along one direction, and AAA crystal with orientational order and positional order along the three spatial directions. Particles in the isotropic and the nematic (the smectic and the crystal) states are colored according to their orientation (vertical coordinate). In the AAA crystal the particles in different layers are located on top of each other whereas in the ABC crystals their location is like in a FCC crystal. Sketches created with OVITO~\cite{Stukowski2009}.
	}
  \label{fig1}
\end{figure*}

In our study of hard spherocylinders, we use the EOS proposed by~\citeau{Peters2020a}, which is depicted in~\cref{fig1} for four different values of the aspect ratio $L/D$, where $L$ is the length of the cylinder and $D$ the diameter of the spherocylinders.
We represent the EOS in the plane of $\mu$ and $\eta$, with $\eta=\rho v_0$ the packing fraction and $v_0$ the particle volume.

\subsection{Construction of the stacking diagram}

The analogue to the bulk phase diagram in presence of a gravitational field is the stacking diagram~\cite{Heras2013} that groups all possible stacking sequences of a given colloidal suspension in sedimentation-diffusion-equilibrium.
To construct the stacking diagram one needs to find the sedimentation binodals that determine the boundaries between distinct stacking sequences.

In monodisperse and mass-polydisperse suspensions, the sedimentation binodals are formed by a set of three different types of sedimentation paths: (i) paths that end at a bulk binodal, \ie $\mue(z=0)=\muc$ with $\muc$ the chemical potential at coexistence in a bulk transition, (ii) paths that start at a bulk binodal, \ie $\mue(h)=\muc$, and (iii) path that are tangent to a bulk binodal. 
Examples of all types of paths are shown in Fig.~\ref{fig0}(a) and Fig.~\ref{fig0}(b) for monodisperse and mass-polydisperse suspensions, respectively.
The third type of paths (tangent to a bulk binodal) is only present in mass-polydisperse suspensions since there the sedimentation path can be curved.
These three types of sedimentation paths form the sedimentation binodals because an infinitesimal change of the path can alter the stacking sequence.

We construct the stacking diagrams in the (experimentally relevant) plane of average colloidal packing fraction of the sample and sample height.
A detailed account of the construction of stacking diagrams in monodisperse and mass-polydisperse systems is given in Ref.~\cite{Eckert2022b}.

\section{Results}

We start in Sec.~\ref{RA} summarizing the main results for the bulk of the system.
The effect of the gravitational field is then presented in Sec.~\ref{RB} (stacking diagrams), Sec.~\ref{RC} (effect of sample height), and Sec.~\ref{RD} (effect of mass-polydispersity).

\subsection{Bulk phase behaviour}\label{RA}
Several theoretical and simulation techniques have been used to study the equation of state of hard spherocylinders.
These include Monte Carlo simulations~\cite{Few1973,VieillardBaron1974}, molecular dynamics simulations~\cite{Rebertus1977}, Brownian dynamics simulations~\cite{Loewen1994,Yan2019}, scaled particle theory~\cite{Lasher1970,Cotter1977} and cell model theory~\cite{Graf1999}.
We use here the EOS by~\citeau{Peters2020a} to model the bulk properties of the system.
These closed-form equations of state are simple yet they provide an accurate description of the full phase behaviour of hard spherocylinders, depicted in Fig.~\ref{fig1}.
We show the bulk phase diagram for rods with four different aspect ratios selected to illustrate the entire phenomenology of hard spherocylinders~\cite{Bolhuis1997,Peters2020a}.
Five different phases occur in bulk, namely isotropic $(I)$, nematic $(N)$, smectic-A $(S)$, AAA crystal $(A),$ and ABC crystal $(C)$.
See schematics in Fig.~\ref{fig1}(e).
For low aspect ratios, \eg $L/D=3.0$ in Fig.~\ref{fig1}(a), there is only a first order phase transition from isotropic fluid to ABC crystal.
Increasing the aspect ratio to \eg $L/D=3.7$ stabilizes also a smectic phase at intermediate densities, see Fig.~\ref{fig1}(b).
The nematic state appears at larger aspect ratios such as $L/D=5$ shown in Fig.~\ref{fig1}(c).
Finally, for rather elongated rods, \eg $L/D=10$ in Fig.~\ref{fig1}(d), an AAA crystal is stable between the smectic and the ABC crystal.
In all cases the phase transitions are of first order (note the jumps in packing fractions at the transitions in Fig.~\ref{fig1}).

\subsection{Stacking diagram of monodisperse spherocylinders}\label{RB}
\begin{figure*}
  \centering
  \includegraphics[width=0.80\linewidth]{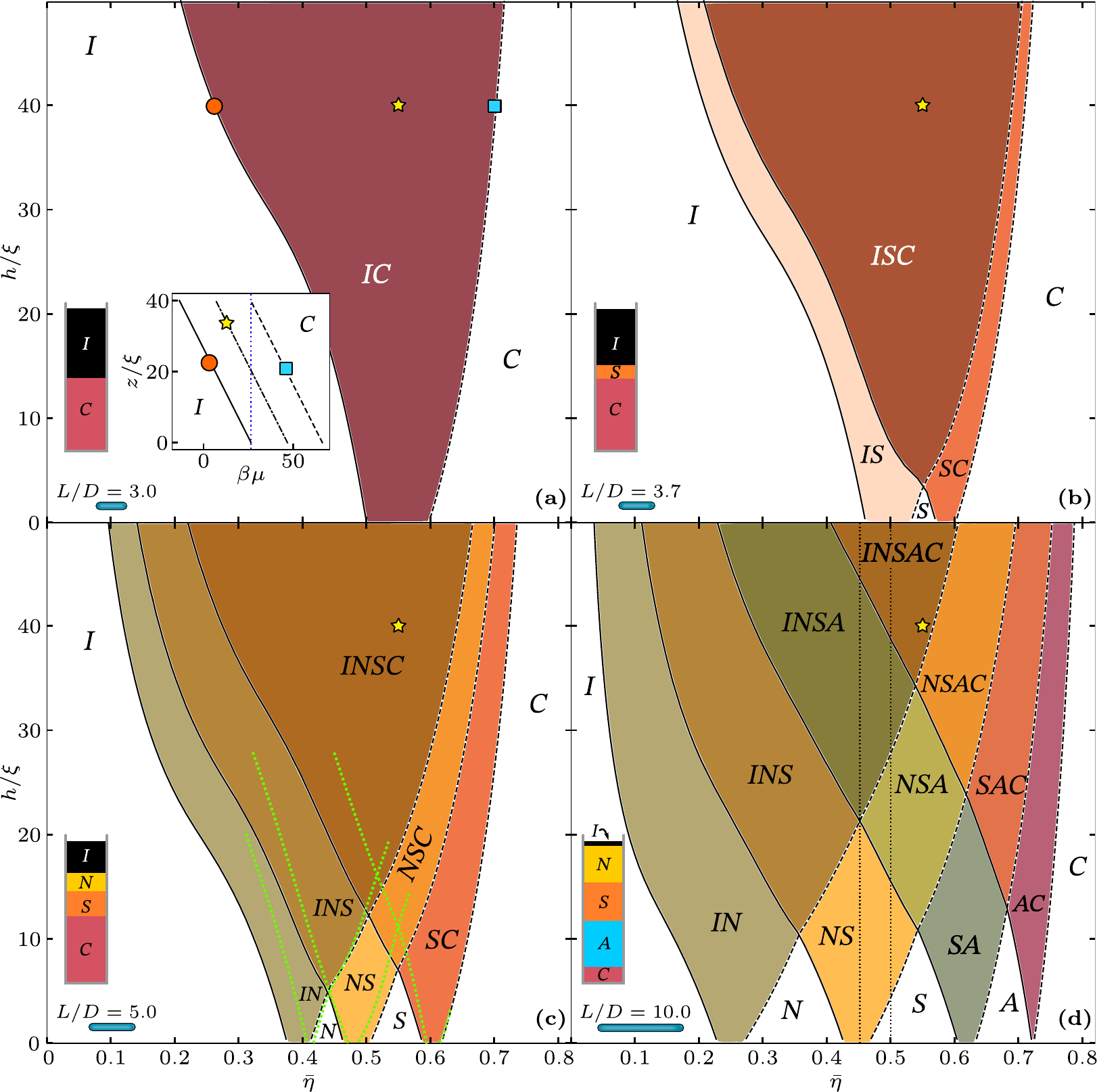}
	\caption{{\bf Sedimentation of monodisperse rods.} Stacking diagram in the plane of average packing fraction $\etab$ and sample height $h/\xi$ for a monodisperse colloidal suspension of hard spherocylinders of different length-to-width aspect ratios: (a) $L/D$=3.0, (b) $L/D$=3.7, (c) $L/D=5.0$, and (d) $L/D=10.0$. Here $\xi$ is the gravitational length.
    The stacking sequences are labeled from top to bottom of the sample.
	The occurring layers are isotropic $(I)$, nematic $(N)$, smectic-A $(S)$, AAA crystal $(A)$, and ABC crystal $(C)$.
	The solid- and dashed-black lines are sedimentation binodals formed by those sedimentation paths that either end or start at a bulk binodal, respectively.
	The dotted green lines in panel (c) are results from~\citeau{Savenko2004}.
    The yellow stars indicate the position of the sample with $(\etab,h/\xi)=(0.55,40)$ in all stacking diagrams.
    A sketch of the corresponding stacking sequences at this state point illustrating the relative thickness of each layer in the stacking sequence is provided within each panel.
    The inset in panel (a) depicts three sedimentation paths in the plane of chemical potential $\mu$ and altitude $z$. The vertical blue line indicates the value  $\beta\muc=26.4$ of the $I$-$C$ bulk transition.
	The path labeled with an orange circle ends at the binodal, \ie $\mu(0)=\muc$, and the path labeled with a blue square starts at the binodal, \ie $\mu(h)=\muc$.
	Both paths form part of the sedimentation binodals in the stacking diagram, as indicated by the orange circle and the blue square in panel (a).
	The path labeled with a star in the inset of panel (a) corresponds to the sketched sample in (a).
  }
  \label{fig2}
\end{figure*}

We next combine the bulk equations of state with sedimentation path theory to obtain the stacking diagrams of hard spherocylinders in a gravitational field.
In~\cref{fig2} we present several stacking diagrams in the plane of average packing fraction $\etab$ and sample height $h$ (scaled with the gravitational length $\xi$) for monodisperse colloidal suspensions of hard spherocylinders with different aspect ratios, as indicated.
In all cases, we assume a positive buoyant mass of the spherocylinders.
Each point in the stacking diagram represents a sedimentation path and therefore a sample under gravity (the sketched samples in Fig.~\ref{fig2} correspond to the points marked by a yellow star in the stacking diagrams).
Depending on the aspect ratio, we find systems which form stacking sequences with up to five different layers.

Since all particles share the same buoyant mass,
the effective local chemical potential $\mue(z)$ is just linear in the vertical coordinate $z$ and there exist only two types of sedimentation binodals.
Those are formed by the set of sedimentation paths that either start (dashed-lines in Fig.~\ref{fig2}) or end (solid-lines in Fig.~\ref{fig2}) at a bulk transition, 
see illustrative examples of sedimentation paths in the inset of Fig.~\ref{fig2}(a).

In the stacking diagram the two types of sedimentation binodals have opposite slope in the plane of $\etab$ and $h$.
Furthermore, two sedimentation binodals of the same type never cross each other and two sedimentation binodals of different type cross each other at most once.
This gives rise to the intertwined pattern of the two types of sedimentation binodals that can be seen in~\cref{fig2}.
When crossing a sedimentation binodal in the stacking diagram, one layer either appears or disappears from either the top or the bottom of the sample.
Four different stacking sequences merge at the points where two sedimentation binodals cross each other. 
At those crossing points, the slope of the sedimentation binodals in the stacking diagram changes as a direct consequence of the density jump of the associated bulk phase transitions (recall that all bulk transition are of first order).
The change in slope is particularly noticeable if the density jump in bulk is large, like \eg for spherocylinders with $L/D=3.7$, cf. Fig.~\ref{fig1}(b) and Fig.~\ref{fig2}(b).
Increasing the sample height also increases the length of the sedimentation path, making it possible to find sequences with more layers.
The maximal number of layers in a monodisperse suspension is always the total number of different bulk phases.
It is worth noting that due to the local equilibrium approximation, the limit $h\rightarrow0$ of the stacking diagram corresponds to 
the bulk of the system. (The sedimentation path becomes a single point in this limit.)

For short spherocylinders, aspect ratio $L/D \lesssim 3.1$, the stable bulk phases are isotropic $(I)$ and ABC crystal $(C)$~\cite{Bolhuis1997}.
Under gravity, the possible stacking sequences are: pure $I$, pure $C$ and $IC$, see~\cref{fig2}(a).
In bulk, \ie in the limit $h\rightarrow 0$ in Fig.~\ref{fig2}, the region of packing fractions where isotropic and columnar phases coexist is $0.5 \le \eta \le 0.6$.
Under gravity, the range of average packing fraction $\etab$ in which we find the stacking sequence $IC$ broadens significantly with increasing sample height.
For example for $h/\xi = 50$ the $IC$ sequence appears in a range of average packing fractions $0.21 < \etab < 0.72$.
This representative example illustrates the importance of including the effect of gravity when analysing and interpreting sedimentation experiments in colloidal science.

For $L/D=3.7$ the stable bulk phases are isotropic $(I)$, smectic-A $(S)$, and ABC crystal $(C)$.
The stacking diagram contains six different stacking sequences, see~\cref{fig2}(b).
Besides the pure $I$, $S$ and $C$ stacks, we also find the sequences $IS$, $SC$ and $ISC$.
Again, the two-phase bulk coexistence regions broaden by increasing the sample height.
However, there are now two regions of phase coexistence in bulk, namely $I+S$ and $S+C$, which give rise to the stacking sequences $IS$ and $SC$.
The regions where $IS$ and $SC$ are stable in the stacking diagram overlap for $h/\xi > 3.3$, forming the additional stacking sequence $ISC$.
Note that there is no isotropic-smectic-crystal triple point in bulk.
The formation of the three layer sequence, $ISC$, is purely due to the gravitational field.
Even tough $I$, $S$, and $C$ phases are stable in bulk, not all possible combinations of layers appear in the stacking diagram.
For example, the occurrence of the sequence $IC$ is not possible for this aspect ratio.
Only two phases that coexist in bulk can appear consecutively in the stacking diagram.

A nematic $(N)$ bulk state is stable for rods with $L/D=5.0$, as depicted in~\cref{fig1}(c).
The corresponding stacking diagram, shown in Fig.~\ref{fig2}(c), contains up to ten different stacking sequences.
In a general monodisperse suspension with no triple points in bulk, the amount of distinct stacking sequences $n_s$ is given by the triangular number of the total number of stable bulk phases $n_b$, that is
\begin{equation}
	n_s=\sum_{i=1}^{n_b}i=\frac{n_b(n_b+1)}2.\label{eq:ns}
\end{equation}

We also show in Fig.~\ref{fig2}(c) results from~\citeau{Savenko2004} (green dotted lines).
They used an EOS from data obtained via computer simulations by \citeau{McGrother1996}.
(For a detailed comparison between the equations of state by \citeauthor{McGrother1996} and by  \citeauthor{Peters2020a} see Ref.~\cite{Peters2020a}.)
\citeau{Savenko2004} used a local equilibrium approximation together with the equilibrium macroscopic condition $\diff P(z)/\diff z=-mg\rho(z)$, with $P$ the osmotic pressure, 
to calculate the sedimentation of hard spherocylinders with $L/D=5.0$.
Their results agree semiquantitatively with our predictions.
The approach by~\citeau{Savenko2004} is equivalent to sedimentation path theory and therefore
the small differences can be attributed to differences in the underlying bulk equation of state (note that both approaches differ slightly in the limit $h\rightarrow0$).
Although both approaches are equivalent, working with the chemical potential is in general more convenient since this quantity varies
linearly with the altitude
within the local equilibrium approximation.
Therefore, the sedimentation path, the density profiles, and the sedimentation binodals can be easily computed.
Moreover, the theory can be straightforwardly applied to binary mixtures~\cite{Heras2013,Geigenfeind2016,Eckert2021} and mass-polydisperse systems~\cite{Eckert2022b}.

By increasing the aspect ratio, hard spherocylinders form an additional crystal AAA ($A$) phase~\cite{Bolhuis1997}, see the bulk diagram for $L/D=10$ in Fig.~\ref{fig1}(d).
This additional bulk phase increases the complexity of the stacking diagram, see~\cref{fig2}(d), which now contains $n_s=15$ different sequences, see Eq.~\eqref{eq:ns}.
Here, the stacking sequence $INSAC$ is stable at large values of sample height $h$ and it contains, in a single sample, all the different bulk phases that hard spherocylinders develop in bulk.
The sample sketched in~\cref{fig2}(d) corresponds to a height $h/\xi=40$. We estimate that in an experimental system made of boehmite rods with polyisobutene coating suspended in
toluene~\cite{Kooij2000} and lengths $L$ between $\SI{1}{\micro m}$ and $\SI{200}{nm}$, the corresponding sample heights would vary between between approx.\ $\SI{0.2}{cm}$  and $\SI{20}{cm}$, respectively.

In all cases, by increasing the sample height, the range of chemical potentials covered by the path becomes larger.
Hence, the sequence with the largest number of layers (\eg $INSAC$ for $L/D=10$) appears always and dominates the stacking diagram for sufficiently large values of the sample height.

We have assumed a positive value of the buoyant mass.
For negative buoyant masses, the only change is a reversed order of the layers in the stacking sequences.
For example, the sequence $ISC$ (from top to bottom) in rods with positive buoyant mass, would be $CSI$ if the rods had negative buoyant mass.

\subsection{Influence of the sample height on the stacking sequences}\label{RC}

\begin{figure}
  \centering
  \includegraphics[width=\linewidth]{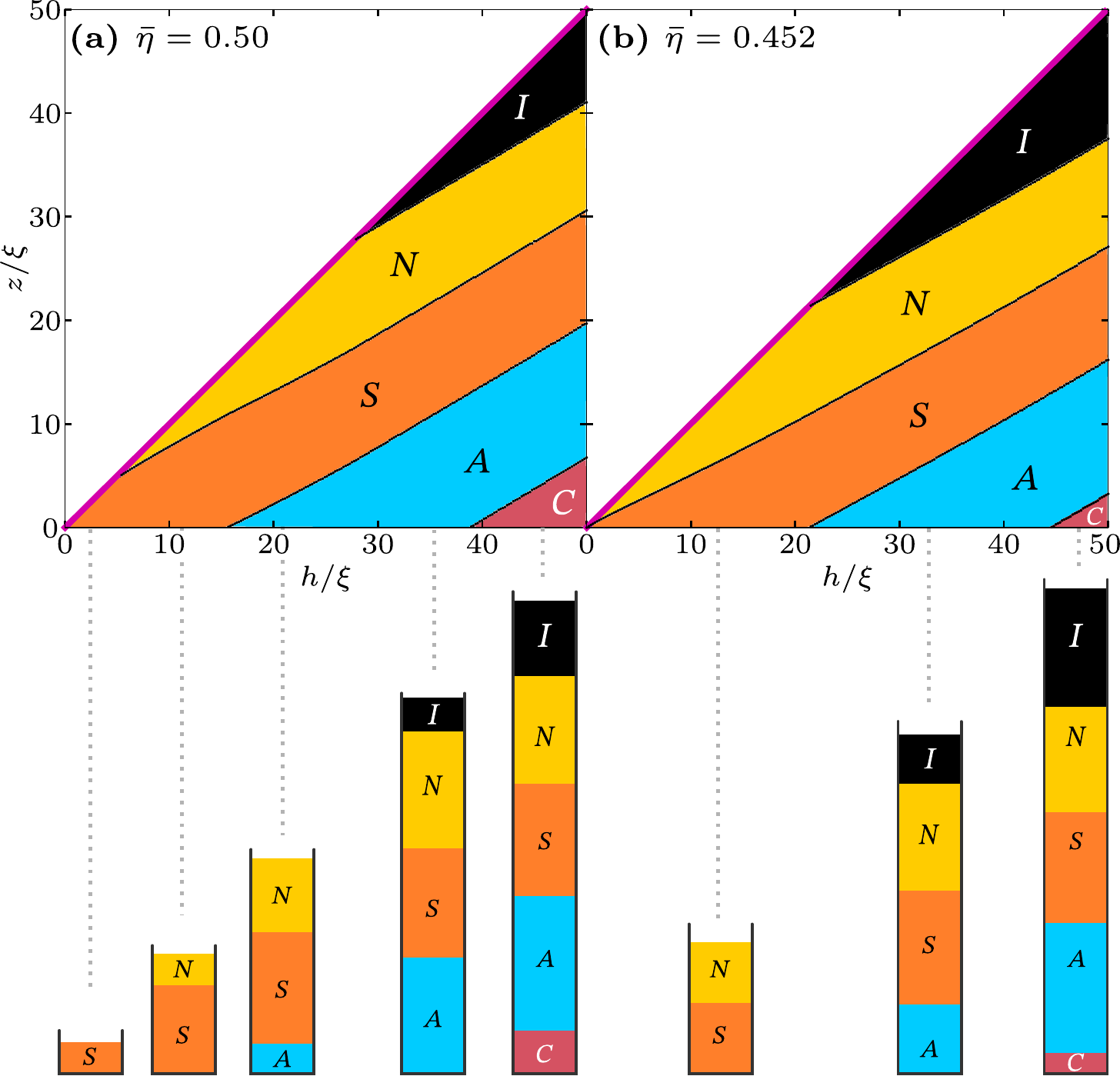}
	\caption{{\bf Effect of sample height.} Stable layer at elevation $z/\xi$ as a function of the sample height $h/\xi$ for colloidal hard spherocylinders with length-to-width aspect ratio $L/D=10.0$.
	The average packing fraction is fixed to (a) $\etab=0.50$ and (b) $\etab=0.452$, indicated by vertical dotted lines in the stacking diagram shown in Fig.~\ref{fig2}(d).
     The layers are isotropic $(I)$, nematic $(N)$, smectic-A $(S)$, AAA crystal $(A)$, and ABC crystal $(C)$.
     The thick violet lines indicate the air-sample interfaces at $z=h$.
	The sketches show all different stacking sequences at selected heights (indicated by vertical dotted lines).
  }
  \label{fig3}
\end{figure}

The effect of varying the sample height (at fixed average colloidal concentration) in sedimentation of colloidal binary mixtures can be counterintuitive with new layers appearing at the top, at the bottom, or in the middle of the sample~\cite{Geigenfeind2016,Eckert2021,Eckert2022}.
We study here the evolution of the stacking sequence by increasing the sample height in monodisperse colloidal systems. 
We select two illustrative average packing fractions $\etab = 0.452$ and $0.50$ in a suspension of elongated hard spherocylinders with $L/D=10.0$, see vertical dotted lines in Fig.~\ref{fig2}(d).
We then vary the sample height while keeping the packing fraction constant and track the stable layer at a given elevation $z$.
The results are shown in~\cref{fig3}.

For $\etab = 0.50$ the system is a pure smectic-A phase in the bulk limit $h \rightarrow 0$, see~\cref{fig3}(a).
By increasing the sample height, a nematic layer develops at the top of the sample, followed by a crystal AAA layer at the bottom.
Next an isotropic layer forms at the top, and finally a crystal ABC layer appears at the bottom of the sample.
There is therefore an alternating pattern of layers growing either at the top or at the bottom of the sample by increasing sample height.

For $\eta = 0.452$, see Fig.~\ref{fig3}(b), the bulk system phase separates into a nematic phase and a smectic-A phase.
Due to the phase separation taking place in bulk, the sedimented samples never show a sequence with a single layer.
Instead, short samples develop the sequence $NS$.
By increasing the sample height two layers develop simultaneously: an isotropic layer is formed on top of the sample and an AAA-crystal forms at the bottom.
The simultaneous introduction of two layers at the top and at the bottom can also be observed in the stacking diagram, see~\cref{fig2}(d).
The dotted vertical line at $\etab = 0.452$ goes through the crossing point between two sedimentation binodals.
Finally, for samples with $h/\xi > 44.3$ an ABC-crystal layer is formed at the bottom of the sample, see~\cref{fig3}(b).
No further changes in the stacking sequence occur for larger values of the sample height.

The evolution of the stacking sequence for two different packing fractions shown in~\cref{fig3} emphasizes the importance of the sample height as a crucial variable in colloidal sedimentation studies~\cite{Geigenfeind2016}.
The stacking diagram results from the bulk diagram.
However, knowing how the stacking sequences evolve by changing the sample height is not obvious from the bulk phase diagram, 
and complex phenomena such as the simultaneous growth of two layers can occur.

\subsection{Effects of mass-polydispersity close to density matching}\label{RD}

\begin{figure}
  \centering
  \includegraphics[width=0.96\linewidth]{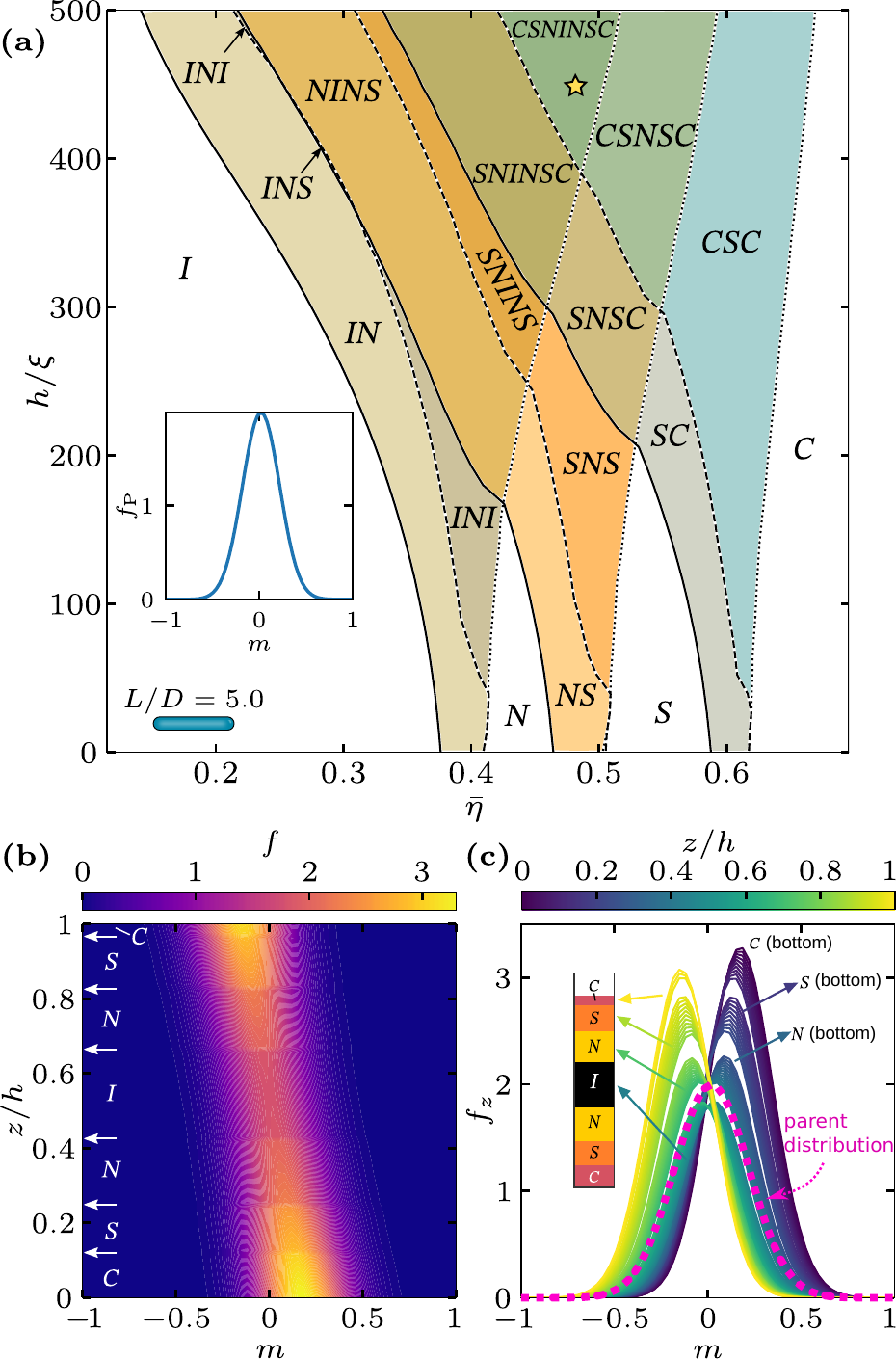}
	\caption{{\bf Sedimentation of mass-polydisperse rods.} (a) Stacking diagram in the plane of average packing fraction $\etab$ and sample height $h/\xi$ of mass-polydisperse colloidal hard spherocylinders with aspect ratio $L/D=5.0$ close to density matching.
    The parent distribution $\fp$ of buoyant masses $m$, a Gaussian with mean 0.02 and standard deviation 0.2, is shown in the inset.
	Here $m=\mbuoy/m_0$ with $m_0$ a reference buoyant mass which corresponds to $5$ times the standard deviation of the parent distribution.
    The stacking sequences are labeled from the top to the bottom of the sample.
	The solid, dashed, and dotted lines are the sedimentation binodals formed by the sedimentation path that either end, start, or are tangent to a bulk binodal, respectively.
	(b) Probability $f(m,z)$ of finding a particle with buoyant mass $m$ at an altitude $z$ in a sample with $h/\xi=450$ and $\etab=0.48$, marked with a yellow star in panel (a).
	The stacking sequence is $CSNINSC$.
	The white arrows indicate the position of the interfaces.
	(c) Probability $f_z(m)$ of finding a particle with mass $m$ at fixed elevation $z$ (see color bar). 
	The initial parent distribution $\fp$ is represented with a dashed-pink line.
	}
  \label{fig4}
\end{figure}

In Ref.~\cite{Eckert2022b} we studied the effects of mass-polydispersity (particles with identical shape but different buoyant masses) on sedimentation.
We identified that mass-polydispersity can play a crucial role in systems that are close to density matching, \ie close to neutral buoyancy.
If the average buoyant mass is close to neutral buoyancy, there will in general be particles in the suspension with positive and negative buoyant masses.
Thus, there exist competing effects with some particles settling under gravity and other particles creaming up.
Close to density matching the effective sedimentation paths are rather horizontal~\cite{Eckert2022b}.
Hence, small changes in the distribution of buoyant masses can have a large effect on the stacking diagram, changing even its topology.

We illustrate here the effects of mass-polydispersity by considering a system of hard spherocylinders with aspect ratio $L/D = 5.0$ and a parent distribution of buoyant masses close to density matching.
The parent distribution gives the probability of finding a particle with buoyant mass $m$ and it is therefore normalized such that $\int \!\diff m\fp(m)=1$.
We use a Gaussian with mean $0.02$ and standard deviation of $0.2$, see inset in~\cref{fig4}(a).
This distribution is close to density matching and it contains particles with positive as well as with negative buoyant masses.
In~\cref{fig4}(a) we show the corresponding stacking diagram in the plane of average packing fraction $\etab$ and scaled sample height $h/\xi$.
The stacking diagram contains $17$ different stacking sequences and it is therefore significantly richer than its monodisperse counterpart ($10$ stacking sequences), cf. Fig.~\ref{fig2}(c) and Fig.~\ref{fig4}(a).
The sedimentation path $\mue(z)$ for mass-polydisperse suspensions is curved and it can have a minimum at intermediate altitudes.
Hence, for each bulk phase transition there is an extra sedimentation binodal formed by those paths for which the minimum of $\mue(z)$ at intermediate altitudes coincides with $\muc$, the value of the chemical potential at bulk coexistence, \ie $\mue(z)=\muc$.
Even though more stacking sequences develop in the mass-polydisperse system than in the monodisperse one, it is worth noting that there seem to exist sequences that only occur in the monodisperse case, \eg $INSC$ and $NSC$.

The sequence with the maximum possible number of layers $CSNINSC$ starts to form in samples with $h/\xi > 395$.
By increasing $h$ this sequence will eventually dominate the stacking diagram and occur for almost any average packing fraction.
All the other stacking sequences that we observe are subsequences of $CSNINSC$.
However, not all subsequences occur.
For example, we observe the sequences $NS$ and $SNS$, but not the sequence $SN$ (top smectic and bottom nematic).
This is due to the asymmetric parent distribution that in this example contains a larger proportion of particles with positive buoyant mass.

On the upper left part of the stacking diagram in~\cref{fig4}(a), we observe a slim region with the sequence $INS$, followed by the a reentrant $INI$ stacking sequence.
Note that the sequence $INI$ already appeared for smaller sample heights and higher packing fractions.
Sedimentation binodals of mass-polydisperse suspensions can cross each other multiple times.
In contrast, we do not observe multiple crossings of the sedimentation binodals in the case of monodisperse particles, see~\cref{fig2}.
Another difference is that in mass-polydisperse systems, new layers can enter the stacking sequence in the middle of the sample and not only at the top or at the bottom.
For example, by increasing the height at constant average packing fraction $\etab = 0.52$, the sequence changes from $S$ to $SNS$, see Fig.~\ref{fig4}(a).
That is, a nematic layer nucleates in the middle of the sample.

Fractionation effects such as \eg the accumulation of short rods in the isotropic phase occur in suspensions of hard spherocylinders with shape polydispersity~\cite{Speranza2002}.
In our mass-polydisperse suspension all particles have the same shape and hence such effects cannot occur by construction.
However, we do observe mass fractionation induced by the gravitational field.
As an illustration, we show in Fig.~\ref{fig4}(b) and~\ref{fig4}(c) how the distribution of buoyant masses changes along the vertical coordinate in a given sample with stacking sequence $CSNINSC$.
There is a strong mass fractionation with particles with positive (negative) buoyant mass concentrating in the bottom (top) of the sample. 

\section{Conclusions}

We have calculated the stacking diagram of monodisperse suspensions of hard spherocylinders with different aspect ratios and studied the effect of varying the sample height.
The stacking diagrams are significantly richer than their bulk counterparts.
Increasing the sample height results in general in complex stacking sequences with several layers of different bulk phases.
For sufficiently large values of the sample height a stacking sequence with layers from all possible bulk phases develops.

Multilayer stacking sequences are often found in sedimentation experiments.
For example, ~\citeau{Kooij1999} found a sequence with five layers in plate-rod colloidal mixtures.
This experimental observation has recently been linked to the occurrence of bulk quintuple (five-phase) coexistence in model plate-rod mixtures~\cite{Opdam2022}. 
The striking bulk phase phenomenon identified in Ref.~\cite{Opdam2022} occurs in absence of a gravitational field. 

From the viewpoint of sedimentation path theory, the occurrence of the experimentally observed~\cite{Kooij1999} sedimentation stacks is a rather direct consequence of the influence of gravity.
As we have shown~\cite{Eckert2021}, systematically analysing sedimentation paths allows to theoretically predict the number as well as the correct ordering of the experimentally observed~\cite{Kooij1999} stacking sequences.
This quantitative treatment involves no need for higher (e.g. quintuple) multiphase bulk coexistence.
While every two adjacent phases in a stacking sequence necessarily involve a corresponding two-phase bulk binodal, non-adjacent layers need not coexist in bulk,
and in general they will not coexist in bulk.
As an illustrative example, already in a one-component system of rods with aspect ratio $L/D = 10$ the stacking sequence $INSAC$ appears in samples of sufficient vertical height, see Fig.~\ref{fig3}(d), even though only two-phase coexistences ($I-N$, $N-S$, $S-A$, and $A-C$) are stable in bulk, see Fig.~\ref{fig1}(d).
Nevertheless, it would be very interesting to study the influence of the presence of a bulk quintuple point on the resulting stacking sequences that occur under gravity.

We have also studied here the effect of mass-polydispersity in a suspension near density matching.
The coupling between mass-polydispersity and the gravitational field increases the number of possible stacking sequences with respect to the monodisperse case.
In a real colloidal suspension with low shape-polydispersity in \eg the particle diameters, the effect of mass-polydispersity might dominate over that of size-polydispersity, especially near density matching.
Nevertheless, size-polydispersity has a strong influence in the phase behaviour of hard spherocylinders~\cite{Bates1998,Speranza2002,Meyer2015,Filippo2023} and will also have an effect in sedimentation.
An extension of sedimentation path theory to incorporate also the effect of shape-polydispersity will be presented in future work.

It is apparent from the stacking diagrams that the sample height plays a role as relevant as the average packing fraction in the determination of the stacking sequence.
Even though varying the sample height should be straightforward experimentally, we are not aware of experimental studies that have systematically considered the effect of the sample height on the phase behaviour of colloidal systems under sedimentation.

Surface and interfacial effects are not taken into account here due to the assumed local equilibrium condition.
Wetting and layering at the bottom of the sample~\cite{Mori2006,Marechal2007,Sandomirski2011} might be specially relevant for low height samples and high packing fractions.
Due to the local equilibrium approximation, the theory predicts that new layers in the stacking sequence start to develop with an infinitesimally small thickness.
The surface tension will in reality prevent layers to be stable until a critical thickness is reached, affecting therefore the position of the sedimentation binodals in the stacking diagram.
A full minimization of a density functional theory for hard spherocylinders could be used to describe surface and interfacial effects.

\section{Acknowledgements}
This work is funded by the Deutsche Forschungsgemeinschaft (DFG, German Research Foundation) under project number 436306241.

\section{Conflicts of Interest}
There are no conflicts to declare.

\end{document}